\documentstyle[12pt,epsf]{article}
\newcommand{\RB}{/\!\!\!\!R}
\newcommand{\sqk}{{\tilde q}}
\newcommand{\stp}{{\tilde t}}
\newcommand{\scharm}{{\tilde c}}

\begin{document}
\begin{flushright}
\begin{tabular}{l}
   OCHA-PP-115
\end{tabular}
\end{flushright}\baselineskip24pt
\begin{center}
{\Large \bf Current constraints on squark production
 scenarios for the excess of high $Q^2$ events at HERA
}\\
\vspace{4mm}
\renewcommand{\thefootnote}{\fnsymbol{footnote}}
{\bf Eri Asakawa, 
     Jun-ichi Kamoshita\footnote[2]{E-mail: kamosita@theory.kek.jp
                                 }
     and Akio Sugamoto}\\
{\it Department of Physics of Ochanomizu University,
     Tokyo 112, Japan}\\
\vspace{4mm}

\renewcommand{\thefootnote}{\arabic{footnote}}

{\bf ABSTRACT}\\
\vspace{4mm}
\begin{minipage}{14.cm}
\baselineskip14pt
   We examine the stop production scenario for the
   anomalous excess of high $Q^2$ events observed at HERA,
   by taking into account the constraints coming from the
   leptoquark search at Tevatron and
   the atomic parity violation experiments.
   This scenario is shown to survive persistently even 
   under the severe constraints from the other experiments.
   In the analysis, the branching ratios $B_{\stp_j}$ of 
   the decay modes $\stp_j\to ed$ $(j=1,2)$ are found to be
   useful parameters, representing a number of unknown parameters
   in the SUSY models.
   By expressing the stop contribution to the cross section 
   of deep inelastic $ep$ scattering
   $\sigma(ep\to eX)$ in terms of $B_{\stp_1}$ and $B_{\stp_2}$,
   the allowed region is successfully identified in
   the parameter space of $B_{\stp_1}$ and $B_{\stp_2}$.
\end{minipage}
\end{center}

\vspace{4mm}
\baselineskip12pt
 An anomalous excess of high $Q^2$ events in 
 the deep inelastic scattering $e^+p\to e^+X$ at HERA
 has been reported by H1 and ZEUS collaborations\cite{H1,ZEUS,HERA97}. 
 The excess is observed for high momentum transfer of positron,
 $Q^2\geq 15000$GeV$^2$, and 
 for relatively high $x$ region, $x\ge 0.25$,
 where $x$ is the Bjorken scaling variable. 
 The measured cross section for the excess events is 0.71pb
 against the standard model(SM) expectation of 0.49pb.
 In addition, the excess seems to be broadly distributed around 
 $200\sim250$GeV for the invariant mass distribution of 
 the positron-jet system. 
 In the following, 
 let $M$ be the invariant mass of positron-jet system
 represented by $M=\sqrt{xs}$, where 
 the center of mass energy of the $e^+p$ system $\sqrt{s}=300$GeV at HERA.

 In order to explain the excess of high $Q^2$ events,
 several scenarios have been proposed\cite{CnLqSt,Cn,LqSt,LQ,St,KK}.
 Among them the most plausible scenarios are 
 the squark production ones in which
 the excess and the $M$ distribution of these high $Q^2$ events
 are caused by the $s$-channel production of the squark 
 following the positron-quark collision.
 Although
 the high $Q^2$ events can be enhanced by a $s$-channel resonance,
 the shape of the $M$ distribution 
 by the $s$-channel production of a single squark is too sharp
 to explain the data.
 Therefore, it is difficult to explain
 the broadness of the $M$ distribution 
 by a single squark production.\footnote{
           The discussion on the single squark production scenario 
           can be translated to that for 
           the scalar leptoquark production scenario with 
           suitable substitution of coupling and mass.}
 Fortunately, in the squark production scenarios,
 two almost degenerate peaks can be arisen
 around $M=200\sim250$GeV,
 and therefore 
 the moderate $M$ distribution may be explained
 due to the poor statistics
 of the current data at HERA experiments\cite{KK}.
 Furthermore,
 although the squark production scenarios are possible to explain
 the features of the excess of high $Q^2$ events at HERA,
 we must take into account the constraints
 coming from both high and low energy experiments.

 In this article, we consider 
 the squark production scenarios in the supersymmetric(SUSY) model
 with $R$-parity breaking($\RB$) interaction 
 and examine whether the squark production scenario
 satisfies the constraints from the current experiments.
 We show that the squark production scenario is consistent with
 the constraints from
 the leptoquark search at Tevatron and
 the measurements of atomic parity violation.
 Also it is pointed out that
 the useful parameter is the branching ratio $B_\sqk$ of 
 the decay mode $\sqk\to eq$,
 when we discuss the consistency of the squark production scenarios
 with the leptoquark search at Tevatron.
 The constraint on the plane of the squark mixing parameter
 $\theta_\sqk$ and the $\RB$ coupling $\lambda'$ 
 is also discussed.

 The existence of the lepton-squark-quark($l$-$\sqk$-$q$)
 couplings is essential to explain 
 the excess and the $M$ distribution of high $Q^2$ events.
 The $l$-$\sqk$-$q$ couplings are induced by
 $\RB$ interaction that results from the trilinear $\RB$
 superpotential.\footnote{ 
           On the other hand, 
           leptoquark is a hypothetical particle that couples to 
           both lepton and quark at the same vertex.} 
 The general form of the $\RB$ superpotential is given by 
 \begin{eqnarray} 
   W_{\not R}= \lambda_{ijk}{\hat L}_i{\hat L}_j{\hat E}^c_k 
       +\lambda'_{ijk}{\hat L}_i{\hat Q}_j{\hat D}^c_k 
       +\lambda''_{ijk}{\hat U}^c_i{\hat D}^c_j{\hat D}^c_k , 
 \label{eqn:GRB} 
 \end{eqnarray} 
 where
 $\hat L$ and $\hat Q$ are the lepton and quark superfields
 in $SU(2)$ doublet representation, whereas
 $\hat E^c$, $\hat U^c$ and $\hat D^c$ are superfields for 
 lepton, up- and down-type quarks in singlet representation,
 respectively; and 
 $i, j$ and $k$ denote the generation indices.
 The $\RB$ interaction Lagrangian 
 relevant to the high $Q^2$ excess events
 is derived from $W_{\RB}$ with $i=k=1$ as follows:
 \begin{eqnarray}
   {\cal L}= \lambda'_{1j1}({\tilde u}_{jL}{\bar d}P_L e
       -\bar{\tilde d}_{R}{\bar e^c} P_L u_j)+h.c  ,
   \label{eqn:RBint}
 \end{eqnarray}
 where $P_L$ is the left handed chiral projection operator.
 In general, the squarks($\sqk_L,\sqk_R$)
 are represented as the mixed states between
 the mass eigenstates ($\sqk_1,\sqk_2$)
 of squark with a mixing parameter $\theta_{\sqk}$
 as follows:
 \begin{eqnarray}
   \sqk_L = \sqk_1\cos {\theta_{\sqk}}
                -\sqk_2\sin {\theta_{\sqk}},\qquad
   \sqk_R = \sqk_1\sin {\theta_{\sqk}}
                +\sqk_2\cos {\theta_{\sqk}}.
 \end{eqnarray}
 As mentioned above, two almost degenerate peaks around $200\sim250$GeV 
 are required to explain the broadness of $M$ distribution.
 The squark mixing is necessary to produce
 two peaks in the $M$ distribution by $\RB$ interaction.

 Hereafter,
 as a representative of the squark production scenarios,
 we consider the stop production scenario, in which case $j=3$, and 
 $\sqk$ should be identified as the stop($\stp$)\ \footnote{
           Scalar charm quark($\scharm$) production scenario
           can be considered when 
           we take $j=2$ and $\sqk=\scharm$\cite{CnLqSt,LqSt,St,KK}.
           The stop production scenario can be paraphrased into
           the scharm production scenario by 
           suitable substitution of mixing angle, coupling and mass.
           Scalar up quark production scenario is, however,
           impossible to explain the event excess at HERA.
           It is because the coupling constant of $\lambda'_{111}$ is 
           already restricted within a very small value 
           by neutrinoless double beta decay\cite{2beta}
           for the scalar up quark mass around 200GeV
           and gluino mass smaller than a few TeV, 
           and as a result its contribution to 
           the excess of high $Q^2$ events at HERA is very small.
             }.
 From the combined data of both H1 and ZEUS,
 the excess events with $Q^2\geq 15000$GeV$^2$ cluster
 around $M\sim200$GeV and
 $M\sim230$GeV.\footnote{
           Since the statistics are not high enough to use 
           H1 and ZEUS data separately,
           we use the combined data of both H1 and ZEUS.
           One may doubt whether the 
           events clustered around $\sim230$GeV can be identified
           as the peak,
           though it may be possible that the 
           events clustered around $\sim200$GeV to be identified
           as the peak.
           In this article,
           we will consider the case that the two peaks exist.
                         }
 The location of the peaks in the $M$ distribution
 should be identified as the mass of the two stops.
 Hereafter we take $m_{\stp_1}=200$GeV and  $m_{\stp_2}=230$GeV.

 There exist many constraints on the 
 absolute value of $\RB$ coupling constants
 or on the products of them\cite{LQlim,RBlim}. Especially,
 the strict constraints come from 
 the proton lifetime\cite{proton}, 
 neutrinoless 2$\beta$ decay\cite{2beta},
 neutrino mass bound($m_{\nu_e}$)\cite{nmass},
 $K^0$-$\bar{K^0}$($D^0$-$\bar{D^0}$ and $B^0$-$\bar{B^0}$) mixing and
 both rare and forbidden decays of mesons\cite{meson},
 $\mu$-$e$ conversion\cite{mue}
 and 
 atomic parity violation(APV)\cite{APV}.

 The non-zero value of $\lambda'_{131}$ is necessary to explain 
 the excess events at HERA by the stop production scenario.
 Hereafter, we suppose only $\lambda'_{131}$ to be non-zero and 
 other $\RB$ coupling constants are to be zero.
 As a result we can relax all the above constraints
 except the one from APV.

\begin{figure}
\epsfxsize = 14 cm
\centerline{\epsfbox{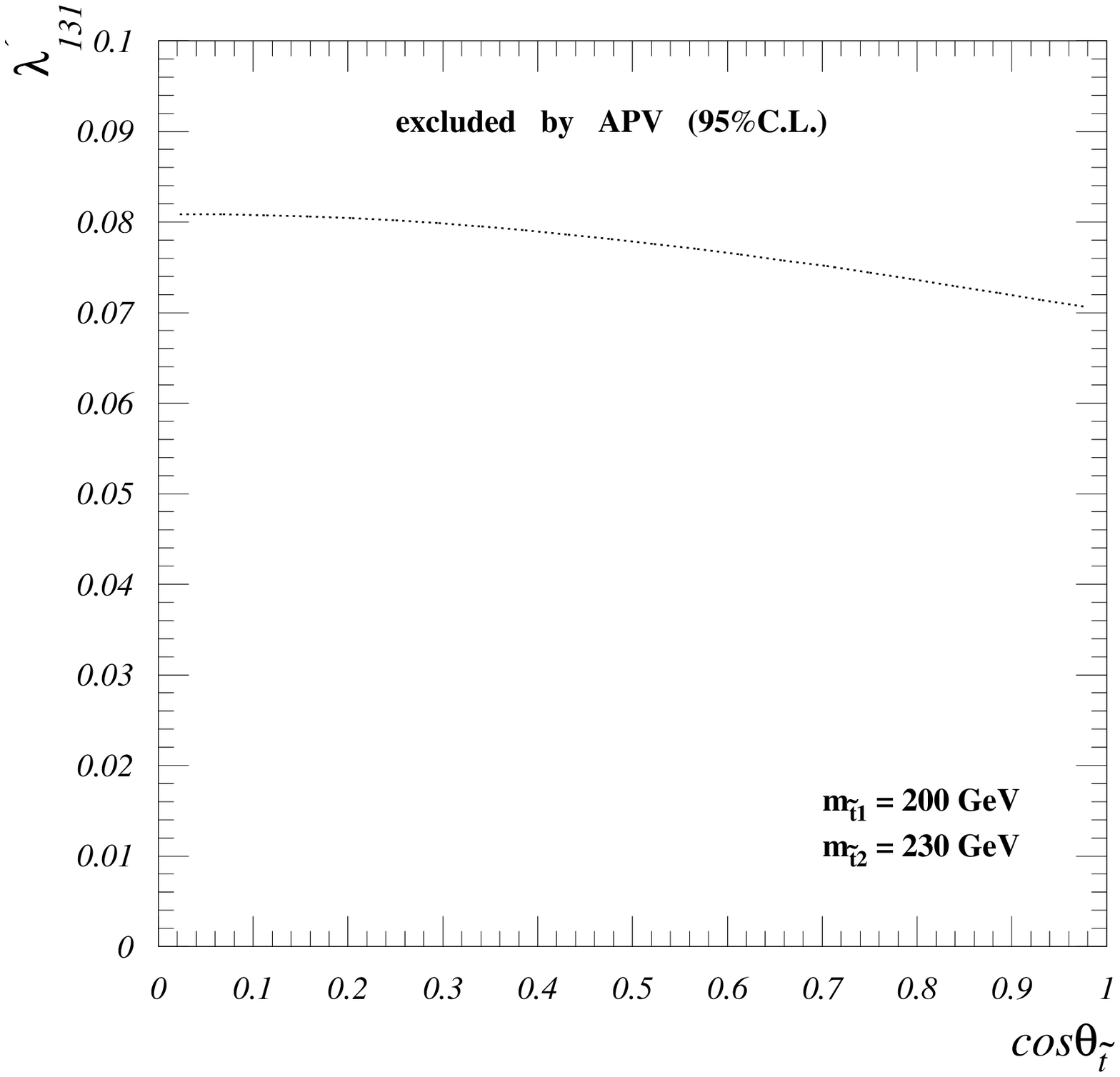}}
\caption[]{ }\label{fig:QW}
\end{figure}

 The most precise measurements of the weak charge $Q^{exp}_W$ 
 in the APV experiments 
 have been obtained from $^{133}$Cs. 
 The weak charge is measured to be 
 $Q^{exp}_W=-72.11\pm0.27\pm0.89$\cite{Wood}. 
 The SM prediction including the radiative correction\cite{Cho} 
 is given as $Q^{SM}_W=-73.11\pm0.05$ leading to 
 $\Delta Q_W\equiv Q^{exp}_W-Q^{SM}_W=1.00\pm0.93$\cite{BCRZ}. 
 The bound on $\Delta Q_W$ with 95\% C.L. is obtained as follows: 
 \begin{eqnarray} 
                   -0.82<\Delta Q_W<2.8. 
 \end{eqnarray} 
 The contribution of the stop to the weak charge\cite{APV} 
 reads 
 \begin{eqnarray} 
     \Delta Q_W=-\frac{|\lambda'_{131}|^2(2N+Z)}{2\sqrt{2}G_F} 
                 \left[ 
                     \frac{\cos^2 \theta_{\tilde t}}{m^2_{\tilde t_1}} 
                   + \frac{\sin^2 \theta_{\tilde t}}{m^2_{\tilde t_2}} 
                \right], 
 \end{eqnarray} 
 where 
 $N=78, Z=55$ for $^{133}$Cs. 
 In figure \ref{fig:QW}, 
 the upper bound on $\lambda'_{131}$ 
 is shown as a function of $\cos\theta_{\tilde t}$. 
 We note from figure \ref{fig:QW} that 
 $\lambda'_{131}\leq 0.07$ is allowed in the  
 whole region of $\theta_{\tilde t}$ 
 when $m_{\tilde t_1}=200$GeV and 
 $m_{\tilde t_2}=230$GeV.\footnote{ 
             The constraints on $\RB$ interaction may be relaxed 
             by the introduction of other new physics sources, 
             for example, the contact interactions\cite{BCRZ}. 
               }

 The formula for the cross section $\sigma(ep\to eX)$ 
 with double stop production can be derived from 
 the straightforward extension of the formula with single stop production 
 in the $\RB$-SUSY model\cite{KK2}. 
 Besides the SM parameters, 
 $\sigma(ep\to eX)$ depends on six variables 
 \begin{equation} 
    \sigma(ep\to eX)=\sigma(m_{\stp_1}, m_{\stp_2}, 
                         \theta_\stp, \lambda'_{131}, 
                         \Gamma_{\stp_1}, \Gamma_{\stp_2}) , 
 \end{equation}
 where $\Gamma_{\stp_j}$ are the total decay widths of $\stp_j$.
 When it is kinematically possible that
 the $\stp_j$ decay modes include SUSY particles, 
 $\Gamma_{\stp_j}$ may depend on a number of
 SUSY parameters.\footnote{
           For example, 
           when the decay mode $\stp\to \chi^0q$ is 
           kinematically allowed,
           the total decay width depends on
           gaugino mass parameters($M_i$ $(i=1,2)$),
           Higgsino mass parameter($\mu$) and 
           the ratio of the vacuum expectation value of 
           two Higgs bosons($\tan\beta\equiv
           \frac{\langle H_2 \rangle}{\langle H_1 \rangle}
           =\frac{v_2}{v_1}$),
           in addition to $m_\stp$, $\theta_\stp$ and $\lambda'_{131}$.
               }
 Once we treat the branching ratios as independent parameters, 
 however,
 the contribution to $\sigma(ep\to eX)$ from the 
 parameters of SUSY model, except for 
 those of the stop sector, can be 
 embedded into the two finite parameters,
  $0\le Br(\stp_j\to ed)\le 1$.

The total decay widths can be expressed as follows: 
 \begin{eqnarray}
      \Gamma_{\stp_j}=
            \frac{\Gamma(\stp_j\to ed)}{Br(\stp_j\to ed)},
 \qquad (j=1,2) ,
 \end{eqnarray} 
 where $\Gamma(\stp_j\to ed)$ are 
 the partial decay widths of the stop to the $ed$ mode, and
 $Br(\stp_j\to ed)$ are the corresponding branching ratios.
 The formulae of $\Gamma(\stp_j\to ed)$ are 
 \begin{eqnarray}
     \Gamma(\stp_1\to ed)=
            \frac{|\lambda'_{131}|^2\cos^2 \theta_{\stp}}{16\pi}
                                            m_{\stp_1}, \quad 
    \Gamma(\stp_2\to ed)=
           \frac{|\lambda'_{131}|^2\sin^2 \theta_{\stp}}{16\pi}
                                            m_{\stp_2}. 
 \end{eqnarray}
 Hereafter we will use shortened notation
 $B_{\stp_j}\equiv Br(\stp_j\to ed)$.
 Then $\sigma(ep\to eX)$ can be rewritten as 
 a function of the six parameters,
 \begin{equation}
    \sigma(ep\to eX)=\sigma(m_{\stp_1}, m_{\stp_2},
                         \theta_\stp, \lambda'_{131},
                         B_{\stp_1}, B_{\stp_2}).
 \end{equation} 
 This choice of parameters is convenient when 
 we discuss the feature of event excess at HERA 
 almost independent of the detailed structure of SUSY model. 
 
 The branching ratios are restricted directly 
 by the first generation scalar leptoquark($LQ_1$) search at 
 Tevatron experiments. 
 The upper bound on the 
 $\sigma(p\bar{p}\to LQ_1\overline{LQ_1}X)B_{LQ_1}^2$ 
 is obtained by the Tevatron experiments\cite{TevLQ},
 where $\sigma(p\bar{p}\to LQ_1\overline{LQ_1}X)$ is the 
 total cross section of the $LQ_1$ pair production and 
 $B_{LQ_1}$ is the branching ratio of the $LQ_1\to eq$ decay mode. 
 The upper bound can be interpreted as that of 
 $\sigma(p\bar{p}\to \stp\bar{\stp}X)B^2_{\stp}$, because 
 one expects that 
 the signal selection cuts for $LQ_1$ can be adopted for the stop
 with $\RB$ interaction(\ref{eqn:RBint}).
 The theoretical prediction of 
 $\sigma(p\bar{p}\to \stp\bar{\stp}X)$ has been estimated
 including SUSY QCD effects
 at the next-to-leading order\cite{PROSTOP}.
 Consequently,
 we can obtain the upper bounds on the branching ratio
 $B_{\stp}$.
 The upper bound are listed in Table \ref{table:Tev}.
 
\begin{figure}
\epsfxsize = 14 cm
\centerline{\epsfbox{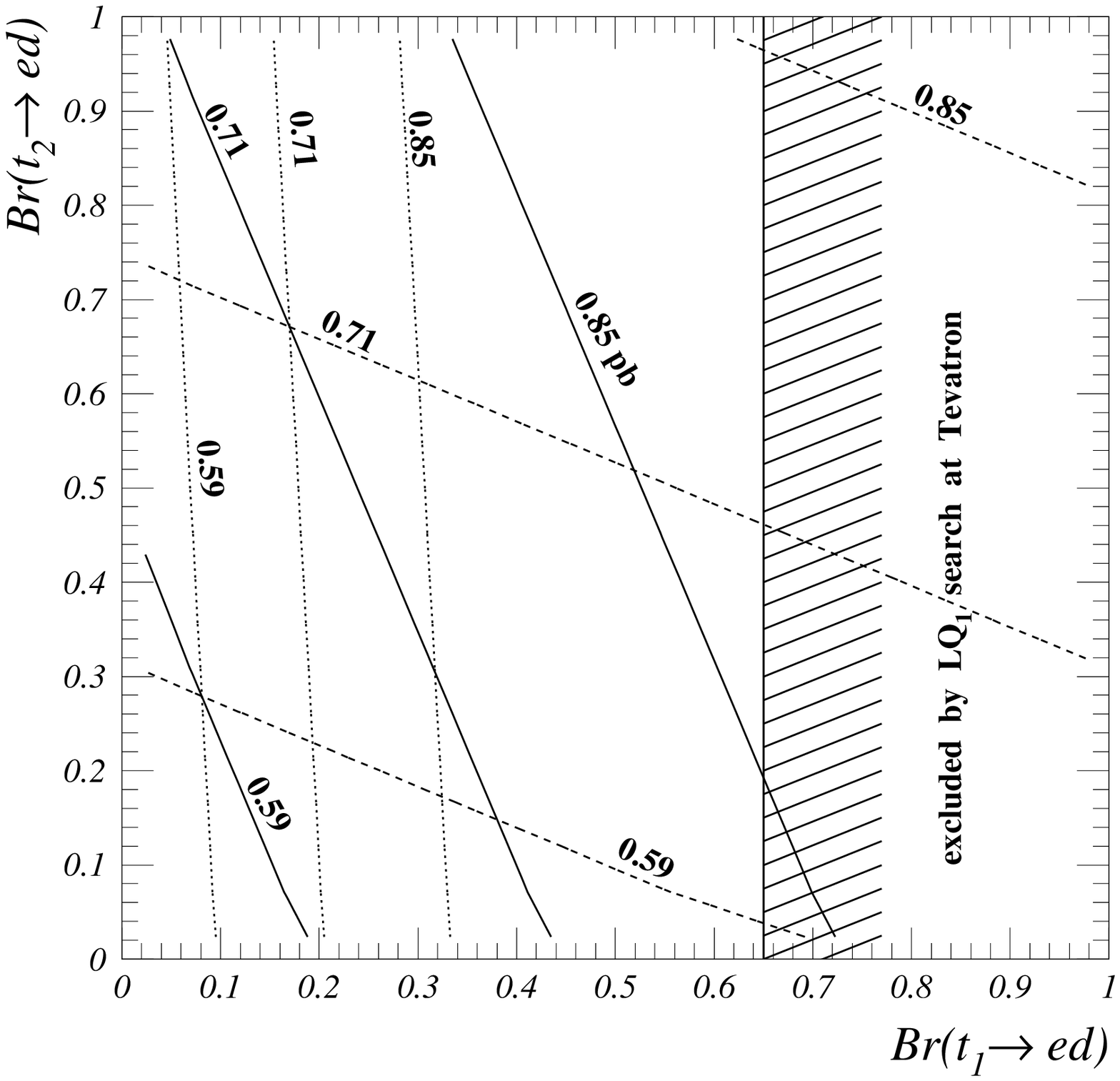}}
\caption[]{ }\label{fig:Br12}
\end{figure}

 Figure \ref{fig:Br12} shows 
 the contour plots of $\sigma(e^+p\to e^+X)$ in the 
 $B_{\stp_1}$ versus $B_{\stp_2}$ plane.
 Although
 the branching ratios are constrained 
 as listed in Table \ref{table:Tev}
 by the first generation leptoquark search at Tevatron,
 there remains a region in the $B_{\stp_1}$-$B_{\stp_2}$ plane
 where the excess of high $Q^2$ events at HERA can be explained
 by the stop production scenario.
 We also find out from figure \ref{fig:Br12} that 
 the contours of the total cross section $\sigma(e^+p\to e^+X)$
 for the same value of $\theta_\stp$ 
 seem to be parallel.
 The reason why the contours are parallel in figure \ref{fig:Br12}
 can be explained as follows.

  The approximate formula of 
 ${\rm d}\sigma(e^+p\to e^+X)/{\rm d}x{\rm d}Q^2$ is obtained
 from its exact formula
 by using the narrow width approximation;
  \begin{eqnarray}
       \frac{{\rm d}\sigma(e^+p\to e^+X)}{{\rm d}x\,{\rm d}Q^2}
   &=&  \left.
        \frac{{\rm d}\sigma(e^+p\to e^+X)}{{\rm d}x\,{\rm d}Q^2}
        \right|_{t-{\rm channel}}
 \nonumber\\
   & & 
        + \frac{\pi\lambda'^2_{131}\cos^2\theta_\stp}{4m^2_{\stp_1}}
                        B_{\stp_1}\ \delta(sx-m^2_{\stp_1})\ d(x,Q^2)
 \nonumber\\
   & &
  + \frac{\pi\lambda'^2_{131}\sin^2\theta_\stp}{4m^2_{\stp_2}}
                        B_{\stp_2}\ \delta(sx-m^2_{\stp_2})\ d(x,Q^2)\ ,
 \label{eqn:dcr}
  \end{eqnarray}
 where the first term on the right hand side,
 denoted by 
  $\left.\frac{{\rm d}\sigma(e^+q\to e^+q)}{{\rm d}x\,{\rm d}Q^2}
        \right|_{t-{\rm channel}}$,
 comes from the $t$-channel exchange of $\gamma, Z$ and 
 $\stp_{j}$ ($j=1,2$),
 whereas the second and third terms come from
 the $s$-channel exchange of $\stp_1$ and $\stp_2$, respectively.
 The parton distribution function for 
 the down quark inside a proton\cite{PDF}
 is denoted as $d(x,Q^2)$.
 Since the amplitude via the $t$-channel exchange of the stop,
 as compared with that by the SM process
 for $m_\stp\ge 200$GeV and $\lambda'_{131}<0.1$,
 is negligibly small,
 the first term on the right hand side in (\ref{eqn:dcr})
 is approximately equal to
 the contribution from the SM.
 Therefore,
 the approximate formula 
 for the total cross section $\sigma(e^+p\to e^+X)$ 
 can be written as follows:
  \begin{eqnarray}
     \sigma(e^+p\to e^+X)&=& 
       \int^1_{x_{min}}\!\!{\rm d}x
       \int^{xs}_{Q^2_{min}}\!\!{\rm d}Q^2
       \frac{{\rm d}\sigma(e^+p\to e^+X)}{{\rm d}x\,{\rm d}Q^2}
 \nonumber\\&=& 
          \left.\sigma(e^+p\to e^+X)\right|_{SM}
 \nonumber\\
    & & + \frac{\pi\lambda'^2_{131}\cos^2\theta_\stp}{4m^2_{\stp_1}}
                   B_{\stp_1}\, I(\frac{m^2_{\stp_1}}{s},Q^2_{min})
       +\frac{\pi\lambda'^2_{131}\sin^2\theta_\stp}{4m^2_{\stp_2}}
                   B_{\stp_2}\, I(\frac{m^2_{\stp_2}}{s},Q^2_{min}) ,
 \nonumber\\
 \label{eqn:dcr2}
  \end{eqnarray}
 where $I(m^2_{\stp_j}/s,Q^2_{min})$ are defined by
  \begin{eqnarray}
     I(m^2_{\stp_j}/s,Q^2_{min})=
       \int^1_{x_{min}}\!\!{\rm d}x\ 
       \int^{xs}_{Q^2_{min}}\!\!{\rm d}Q^2\ 
                d(x,Q^2)\,\delta(sx-m^2_{\stp_j})
   \hspace{5mm}(j=1,2),\label{eqn:IxQ1}
  \end{eqnarray}
 $x_{min}=Q^2_{min}/s$, and
 $Q^2_{min}$ is the selection cut for the high $Q^2$ events.
 Equation(\ref{eqn:dcr2}) shows that $\sigma(e^+p\to e^+X)$
 depends on $B_{\stp_j}$ $(j=1,2)$ linearly.
 This explains the feature of figure \ref{fig:Br12}
 that the contours of the total cross section are almost parallel
 in the $B_{\stp_1}$-$B_{\stp_2}$ plane.

\begin{figure}
\epsfxsize = 14 cm
\centerline{\epsfbox{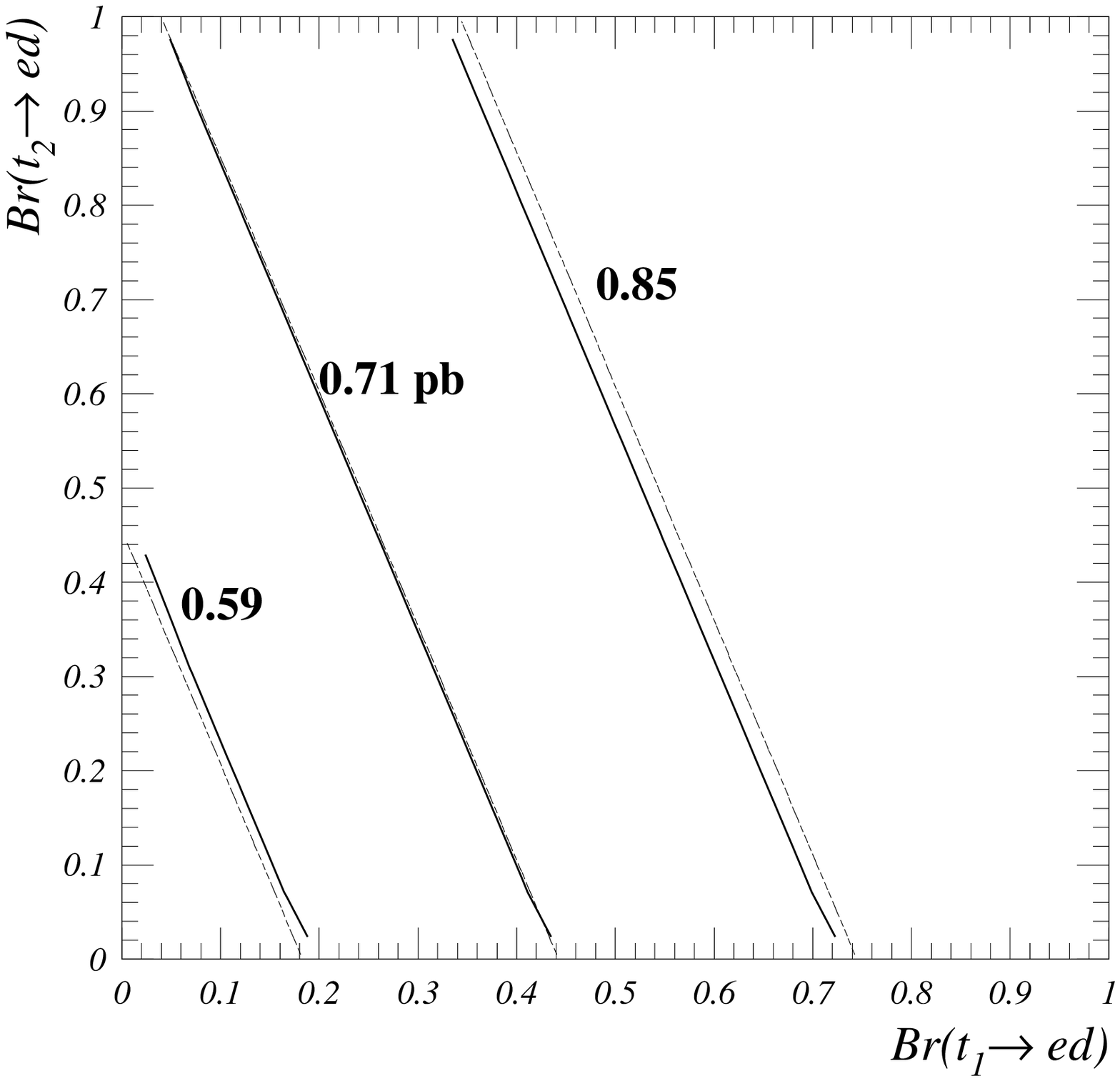}}
\caption[]{ }\label{fig:apprx}
\end{figure}

  After integrating over $x$, we have
  \begin{eqnarray}
     I(m^2_{\stp_j}/s,Q^2_{min})=\left\{\begin{array}{lc}
       \left.\frac{\displaystyle 1}{\displaystyle s}
       { \displaystyle \int^{m^2_{\stp_j}}_{Q^2_{min}} }{\rm d}Q^2\  
          d(x,Q^2)\right|_{x= m^2_{\stp_j}/s}
         & ({\rm if }\ m^2_{\stp_j}\ge Q^2_{min}) \cr
         & \cr
       0 & ({\rm if }\ m^2_{\stp_j}\le Q^2_{min})
       \end{array}\right.
       \hspace{3mm} (j=1,2).\label{eqn:IxQ2}
  \end{eqnarray} 
 Several explicit values of $I(m^2_{\stp_j}/s, Q^2_{min})$ are given
 in Table \ref{table:IxQ}. 
 From eq.(\ref{eqn:dcr2}) and Table \ref{table:IxQ}, 
 we can easily calculate $\sigma(e^+p\to e^+X)$
 for arbitrary values of $\lambda'_{131}$, $\theta_{\stp}$,
 $B_{\stp_1}$ and $B_{\stp_2}$ by substituting
 the two values for $j=1, 2$ of $I(m^2_{\stp_j}/s, Q^2_{min})$
 with the same $Q^2_{min}$.
 In figure \ref{fig:apprx},
 we show the contour plots of $\sigma(e^+p\to e^+X)$
 derived by using the approximate formula(dashed lines) 
 and the exact formula(solid lines).

\begin{figure}
\epsfxsize = 14 cm
\centerline{\epsfbox{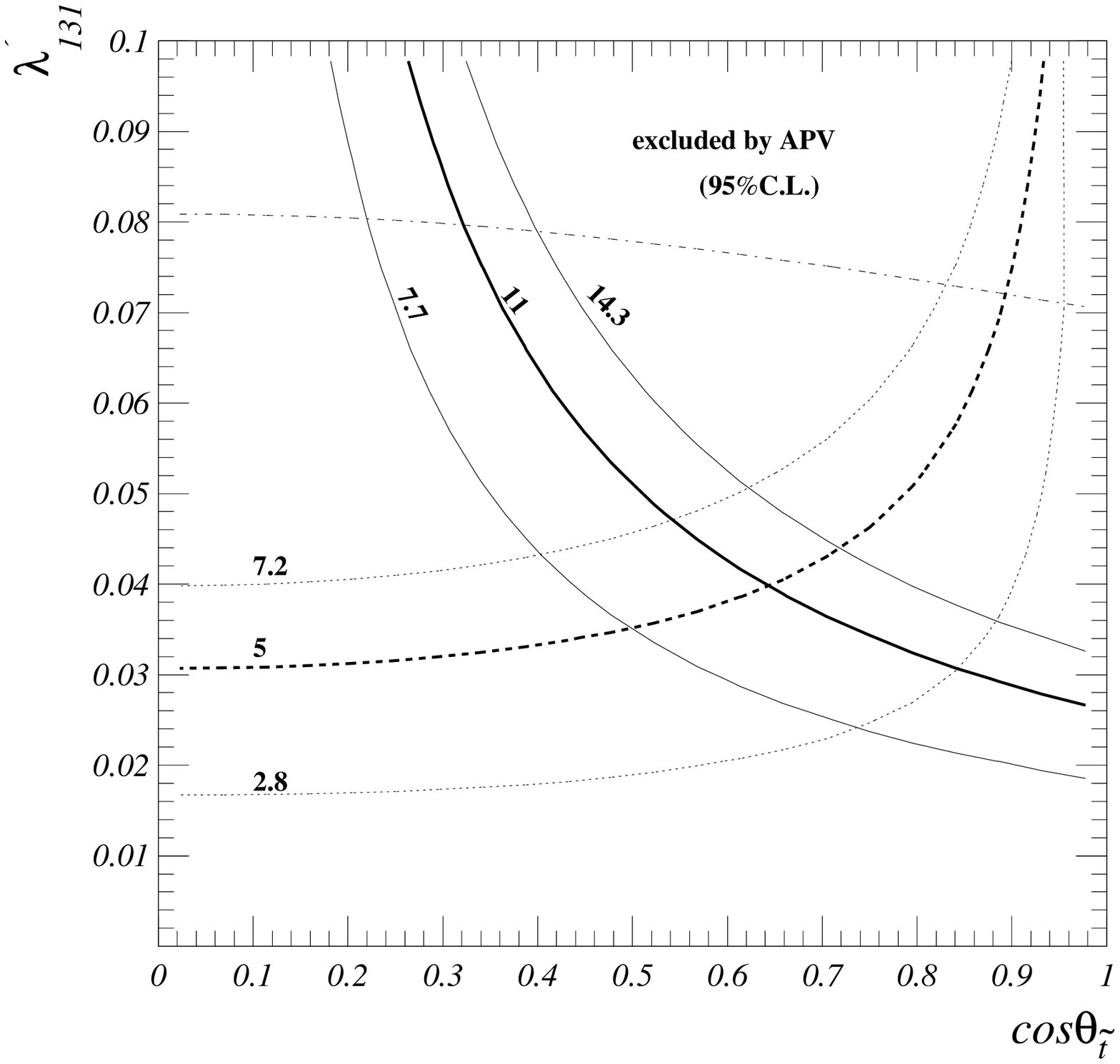}}
\caption[]{(a)}\label{fig:lamb}
\end{figure}

 Next, we discuss whether the constraint on 
 the  $\theta_{\stp}$ and $\lambda'_{131}$ can be obtained
 from the data of H1 and ZEUS for excess events.
 Combining data of H1 and ZEUS
 for $Q^2\geq 15000$GeV$^2$, we can find
 11 events in the first bin 187.5GeV$<M<212.5$GeV and 
  5 events in the second bin 222.5GeV$<M<247.5$GeV\
 with luminosity of $L=57.2$~pb$^{-1}$\cite{HERA97}
 against the SM expectation of
 5.9 events in the first 187.5GeV$<M<212.5$GeV and 
 2.1 events in the second 222.5GeV$<M<247.5$GeV.
 We now require the constraint on the number of events
 in each bin at $1\sigma$ level as follows:
 \begin{eqnarray}
     7.7 \le N|_{187.5{\rm GeV}<M<212.5{\rm GeV}}\le 14.3 ,\qquad
     2.8 \le N|_{222.5{\rm GeV}<M<247.5{\rm GeV}}\le 7.2 .
 \label{eqn:nbin}
 \end{eqnarray}
 Figures\ref{fig:lamb}(a) and \ref{fig:lamb}(b) show
 the contour plots of the expected number of events
 for several sets of $B_{\stp_1}$ and $B_{\stp_2}$,
 along with the upper bound on $\lambda'_{131}$ 
 obtained by APV measurements.\footnote{
          In the calculation of the expected number of events
          in the $\RB$-SUSY model,
          we use the integrated luminosity $L=57.2$pb$^{-1}$ 
          set the efficiency to 1.
          If the efficiency is set to 0.9,
          the contours in figures\ref{fig:lamb}(a)(b)
          should be shifted somewhat above. 
          However the results are not changed significantly.
              }
 In figures\ref{fig:lamb}(a) and \ref{fig:lamb}(b),
 the constraint on the $\cos\theta_{\stp}$-$\lambda'_{131}$ plane
 becomes more strict for smaller values of $B_{\stp_j}$.
 It is because the total decay width of $\stp$ becomes larger 
 for smaller values of $B_{\stp_j}$, and as a result 
 the total cross section of $\sigma(e^+p\to e^+X)$ decreases
 with decreasing $B_\stp$ for a fixed value of $\lambda'_{131}$.
 Therefore, to explain the excess of events,
 we must take the larger values of $\lambda'_{131}$
 corresponding to the smaller values of $B_{\stp_j}$ $(j=1,2)$.
 The value of $\lambda'_{131}$, however, cannot be taken 
 too large for the small $B_{\stp}$,
 since $\lambda'_{131}$ larger than 0.07-0.08 is 
 excluded by APV.

 To conclude our discussion,
 we show that the stop production scenario
 for the excess of high $Q^2$ events at HERA
 is consistent with the current experiments,
 even though the leptoquark search at Tevatron and APV measurements
 strictly constrain the stop production scenario.

 Currently, 
 the statistics at HERA experiments are not enough
 to constrain the $\cos\theta_{\stp}$-$\lambda'_{131}$
 parameter space strictly.
 In the future high luminosity run at HERA,
 we expect that 
 the number of events in each bin 
 will be restricted within a narrower region than 
 the present one in eq.(\ref{eqn:nbin}), 
 so that a more strict constraint on 
 the $\cos\theta_{\stp}$-$\lambda'_{131}$ parameter space 
 will be obtained in the near future. 
 HERA upgrade is planned with 
 $L\approx7.4\times 10^{-5}$pb$^{-1}$s$^{-1}$ 
 starting in the year
 2000\cite{HERA}.\footnote{ 
            The current accumulation of luminosity performance 
            is $L\approx10^{-5}$pb$^{-1}$s$^{-1}$. 
                           }
 Furthermore, 
 if a large amount of luminosity is accumulated 
 at the future Tevatron experiment such as 
 TEV33($L=210.62$pb$^{-1}/$~week)\cite{TEV33}
 by the end of year 2006, 
 then we may expect that 
 the branching ratios $B_{\stp_j}$ ($j=1,2$) will be restricted 
 within the small values, 
 namely $B_{\stp_1}<0.2$ and  $B_{\stp_2}<0.33$. 
 Therefore the stop production scenario will be strictly 
 constrained by TEV33. 
 Nevertheless,
 the figures \ref{fig:Br12} and \ref{fig:lamb} show that 
 the stop production scenario can be compatible with
 the constraints from both 
 the leptoquark search at TEV33 and the measurements of APV, 
 even if the strict constraint on the $B_{\stp_j}$ ($j=1,2$) is 
 obtained at TEV33.

  The authors thanks M. Ahmady 
  for reading the manuscript and helpful comments.
  One of the authors
 (J. K.) thanks T. Kon and F. Shibata for helpful discussion and comments.

\bigskip

\newpage
\begin{center}
  {\Large   Figure captions}
\end{center}

\begin{description}
   \item[Figure \ref{fig:QW}] The upper bound on the value of
          $\lambda'_{131}$ from the measurements of the atomic 
          parity violation(APV) is shown as a function of 
          $\cos\theta_{\stp}$.
          We take ($m_{\stp_1}$, $m_{\stp_2}$)=(200GeV, 230GeV).
   \item[Figure \ref{fig:Br12}] Contour plots of the cross
          section of $\sigma(e^+p\to e^+X)$ 
          for $Q^2\geq 15000$GeV$^2$ are shown in the 
          $B_{\stp_1}$ versus $B_{\stp_2}$ plane;
          the dashed line is for $\cos\theta_{\stp}=0.3$,
          the solid line for $\cos\theta_{\stp}=0.6$ and
          the dotted line for $\cos\theta_{\stp}=0.9$.
          We take $\lambda'_{131}=0.07$ and $\sqrt{s}=300$GeV.
          Each value written beside the curve is corresponds to 
          the measured value,
          $\sigma(e^+p\to e^+X)=0.71^{+0.14}_{-0.12}$pb,
          for $Q^2\geq 15000$GeV$^2$.
   \item[Figure \ref{fig:apprx}] Contour plots of the cross
          section of $\sigma(e^+p\to e^+X)$ 
          for $Q^2\geq 15000$GeV$^2$ and $\cos\theta_{\stp}=0.6$
          are shown in the $B_{\stp_1}$ versus $B_{\stp_2}$ plane;
          the dashed and solid lines are corresponds to
          the results obtained by the approximate formula and
          the exact formula, respectively.
          Other input parameters are the same as
          in figure \ref{fig:Br12}.
   \item[Figure \ref{fig:lamb}(a)(b)] 
          Contour plots of the expected number of events in 
          each bin;  187.5GeV$\le M\le$212.5GeV 
               and   222.5GeV$\le M\le$247.5GeV;
          are shown in the $\cos\theta_{\stp}$-$\lambda'_{131}$ plane,
          where $\sqrt{s}=300$GeV.
          As for the branching ratio of $\stp_j$ $(j=1,2)$,
          we take (a):($B_{\stp_1}$, $B_{\stp_2}$)=(0.65, 1.0)
             and  (b):($B_{\stp_1}$, $B_{\stp_2}$)=(0.2, 0.33).
          The integrated luminosity is taken to be 57.2pb$^{-1}$.
          The curve of the upper limit on $\lambda'_{131}$ by APV
          are also superposed.
\end{description}

\newpage
\baselineskip24pt
\begin{center}
  {\Large   Tables}
\end{center}

 \begin{table}[h]
 \begin{center}
 \begin{tabular}{c|c|c|c} 
           & experimental   &  experimental  & theoretical \\
  mass     & upper bound on & upper bound on & prediction\cite{PROSTOP}  \\
           & & & \\ 
 $m_\stp$ & $Br(\stp\to ed)$
   & $\sigma(p\bar{p}\to \stp\stp X)\{Br(\stp\to ed)\}^2$ 
   & $\sigma(p\bar{p}\to \stp\stp X)_{theor.}$\\
 \hline
  200 GeV  &$\le 0.65 $ & $\le$ 0.076 pb & 0.18 pb\\
  230 GeV  &$\le 1.0 ({\bf no\ bound})  $ & $\le$ 0.067 pb & 0.065 pb\\
 \end{tabular}
 \end{center}
 \begin{center}
 \begin{minipage}{13cm}\caption[]{\small
     The upper bound on $Br(\stp\to ed)$ and
    $\sigma(p\bar{p}\to \stp\stp X)\{Br(\stp\to ed)\}^2$
    from the first generation leptoquark search at Tevatron\cite{TevLQ}.
            }\label{table:Tev}
 \end{minipage}
 \end{center}
 \end{table}

 \begin{table}[h]
 \begin{center}
 \begin{tabular}{c|cc|c} 

 $m_\stp$ & $Q^2_{min}=10000$GeV$^2$ & & $Q^2_{min}=15000$GeV$^2$\\
          &                          & &                         \\
 \hline
  180 GeV  &  7.061\ 10$^-2$    &  &  5.438\ 10$^-2$ \\
  190 GeV  &  5.561\ 10$^-2$    &  &  4.456\ 10$^-2$ \\
  200 GeV  &  4.160\ 10$^-2$    &  &  3.434\ 10$^-2$ \\
  210 GeV  &  2.937\ 10$^-2$    &  &  2.481\ 10$^-2$ \\
  220 GeV  &  1.937\ 10$^-2$    &  &  1.668\ 10$^-2$ \\
  230 GeV  &  1.177\ 10$^-2$    &  &  1.028\ 10$^-2$ \\
  240 GeV  &  6.437\ 10$^-3$    &  &  5.694\ 10$^-3$
 \end{tabular}
 \end{center}
 \begin{center}
 \begin{minipage}{13cm}\caption[]{\small
      Values of $I(m_\stp/s, Q^2_{min})$ for 
      several values of the stop mass($m_\stp$) and $Q^2_{min}$.
      Definition of $I(m_\stp/s, Q^2_{min})$ is given by 
     eqs.(\ref{eqn:IxQ1}) and (\ref{eqn:IxQ2}).
            }\label{table:IxQ}
 \end{minipage}
 \end{center}
 \end{table}

\newpage

%
%

\addtocounter{figure}{-1}

\begin{figure}
\epsfxsize = 14 cm
\centerline{\epsfbox{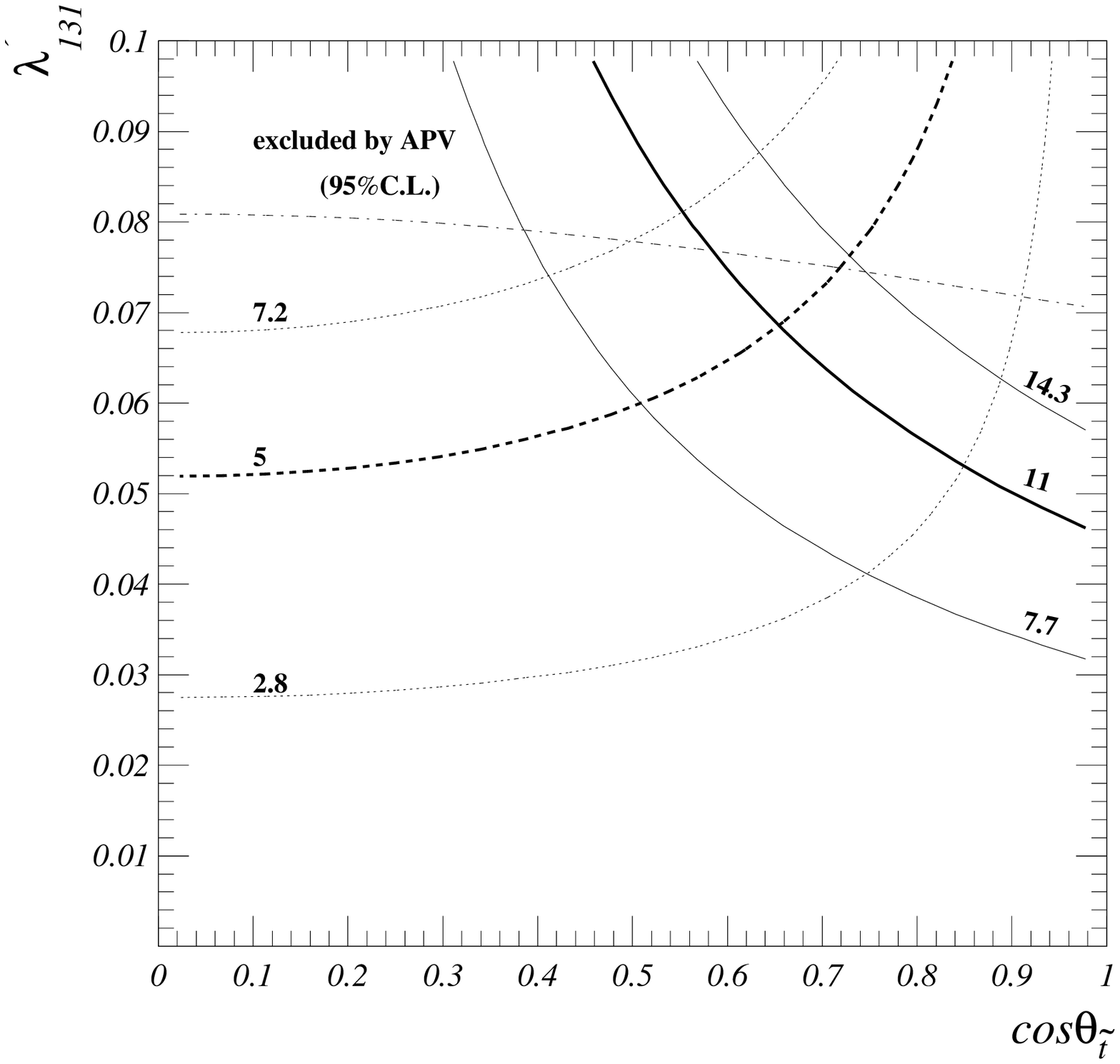}}
\caption[]{(b)}
\end{figure}

\end{document}